\documentclass{egpubl} 
 
\usepackage[T1]{fontenc}
\usepackage{dfadobe}   
\usepackage{cite} 
\usepackage{hyperref}
\BibtexOrBiblatex

\ifpdf \usepackage[pdftex]{graphicx} \pdfcompresslevel=9
\else \usepackage[dvips]{graphicx} \fi
\usepackage{egweblnk} 

\usepackage{adjustbox}
\usepackage{multirow}
\usepackage{amssymb}
\usepackage{color}
\usepackage{tikz}
\usepackage{pgfplotstable}
\usepackage{pgfplots}
\usepackage[normalem]{ulem}  
\usepackage{pifont} 
\usepackage{lipsum}
\usepackage{xspace}
\usepackage{multirow}
\usepackage{amsmath}

\usepackage[table]{xcolor}
\definecolor{best}{rgb}{1.0, 0.749, 0.749}
\definecolor{bestt}{rgb}{0.996, 0.898, 0.792} 
\definecolor{orange2}{rgb}{0.937, 0.553, 0.294}
\usepackage{makecell}
\usepackage{arydshln} 
\usepackage{orcidlink}
\usepackage{cleveref} 
\newcommand{\mysystem}{{GS-Share}\xspace}
\newcommand{\mydataset}{{Replica-Share}\xspace}

\title[GS-Share]{GS-Share: Enabling High-fidelity Map Sharing with \\ Incremental Gaussian Splatting }

\author[Zhang et al.]   
{  
    Xinran Zhang$^{1}$\orcidlink{0009-0002-1589-4222} 
    \quad Hanqi Zhu$^{1}$\orcidlink{0000-0001-5315-1839} 
    \quad Yifan Duan$^{1}$\orcidlink{0009-0004-0754-3953} 
    \quad Yanyong Zhang$^{1,2}$\thanks{Corresponding author.}\orcidlink{0000-0001-9046-798X} 
    \\
\parbox{\textwidth}{\centering  
         $^1$ University of Science and Technology of China, China \\
         $^2$ Institute of Artificial Intelligence, Hefei Comprehensive National Science Center, China 
       }
}

\begin{document}
 
\teaser{
 \includegraphics[width=0.9\linewidth]{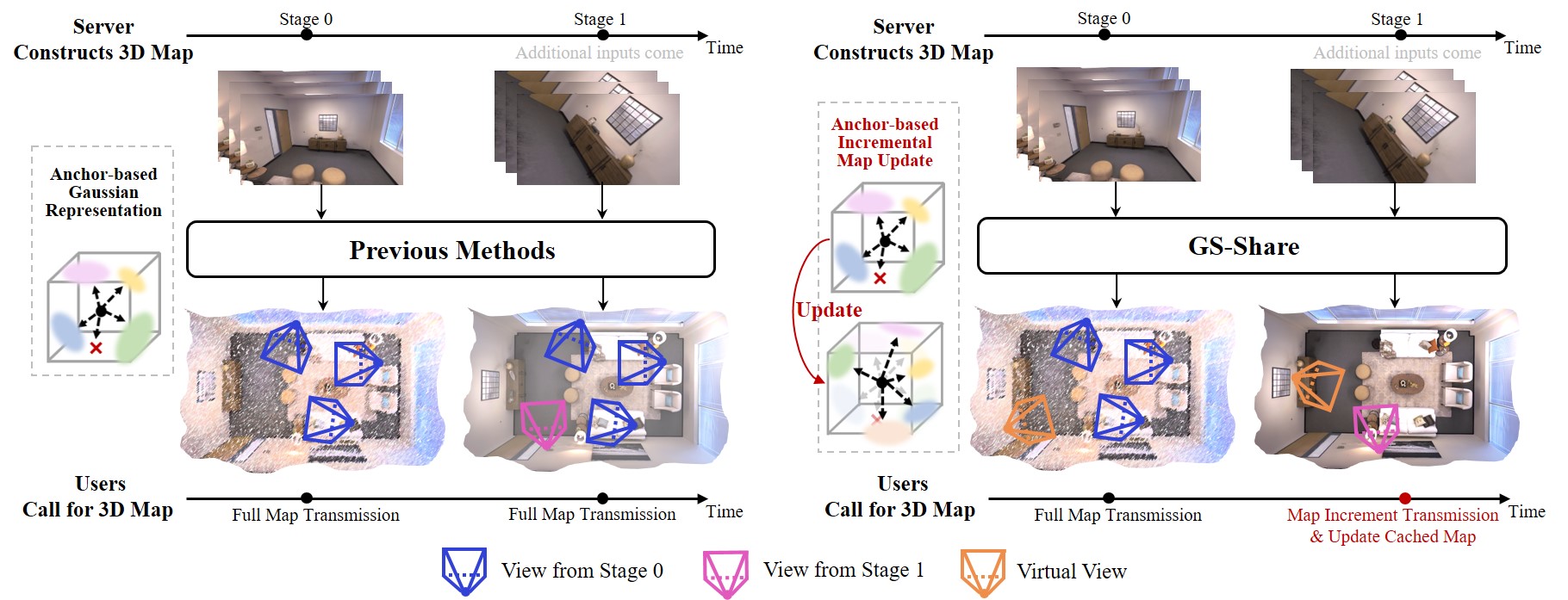}
 \centering
  \caption{Comparison between previous methods and our GS-Share. In map-sharing scenarios where input images arrive progressively, previous methods typically reconstruct the entire map from scratch at each stage, requiring full transmission whenever a user updates their map. In contrast, GS-Share adopts an incremental map update strategy that learns only the changes and transmits only the map increments, effectively eliminating redundant data transfer. Beyond reducing transmission overhead, GS-Share also leverages virtual-view synthesis to enhance mapping accuracy. Together, these techniques form a complete map-sharing framework tailored to the Gaussian representation.
  }
\label{fig:first}
}

\maketitle

\begin{abstract}  
Constructing and sharing 3D maps is essential for many applications, including autonomous driving and augmented reality. Recently, 3D Gaussian splatting has emerged as a promising approach for accurate 3D reconstruction. However, a practical map-sharing system that features high-fidelity, continuous updates, and network efficiency remains elusive. To address these challenges, we introduce \mysystem, a photorealistic map-sharing system with a compact representation. The core of \mysystem includes anchor-based global map construction, virtual-image-based map enhancement, and incremental map update.    
We evaluate \mysystem against state-of-the-art methods, demonstrating that our system achieves higher fidelity, particularly for extrapolated views, with improvements of 11\%, 22\%, and 74\% in PSNR, LPIPS, and Depth L1, respectively. Furthermore, \mysystem is significantly more compact, reducing map transmission overhead by 36\%. 


\end{abstract}

\section{Introduction}\label{sec:intro}
Exploring and mapping uncharted environments has long been an enduring pursuit of humanity. Imagine a bustling shopping mall where a high-fidelity 3D map can seamlessly guide visitors to their destinations, offering dense and detailed reconstructions that significantly enhance user experiences.
To realize this vision, 3D Gaussian Splatting (3DGS)~\cite{3DGS} has emerged as one of the most promising approaches. By explicitly modeling each element in the map as a Gaussian ellipsoid, 3DGS achieves remarkable results.
However, utilizing 3DGS at the scene level, as opposed to the object level, typically involves massive data requirements.
Consequently, crowd-sourcing is typically employed for data collection, enabling dispersed participants to share their map data~\cite{crowdsoursing}, which is managed through a map-sharing framework. 
Typically, such a map-sharing framework involves two categories of participants~\cite{map++}: contributors and users. Contributors are individuals who traverse previously uncharted areas and upload their observations to the server. In contrast, as shown in Fig.~\ref{fig:first}, users download the map from the server and convert it into a more human-readable format, such as rendering it into images.

\begin{figure}[t]
	\centering
	\includegraphics[width = 0.97\linewidth]{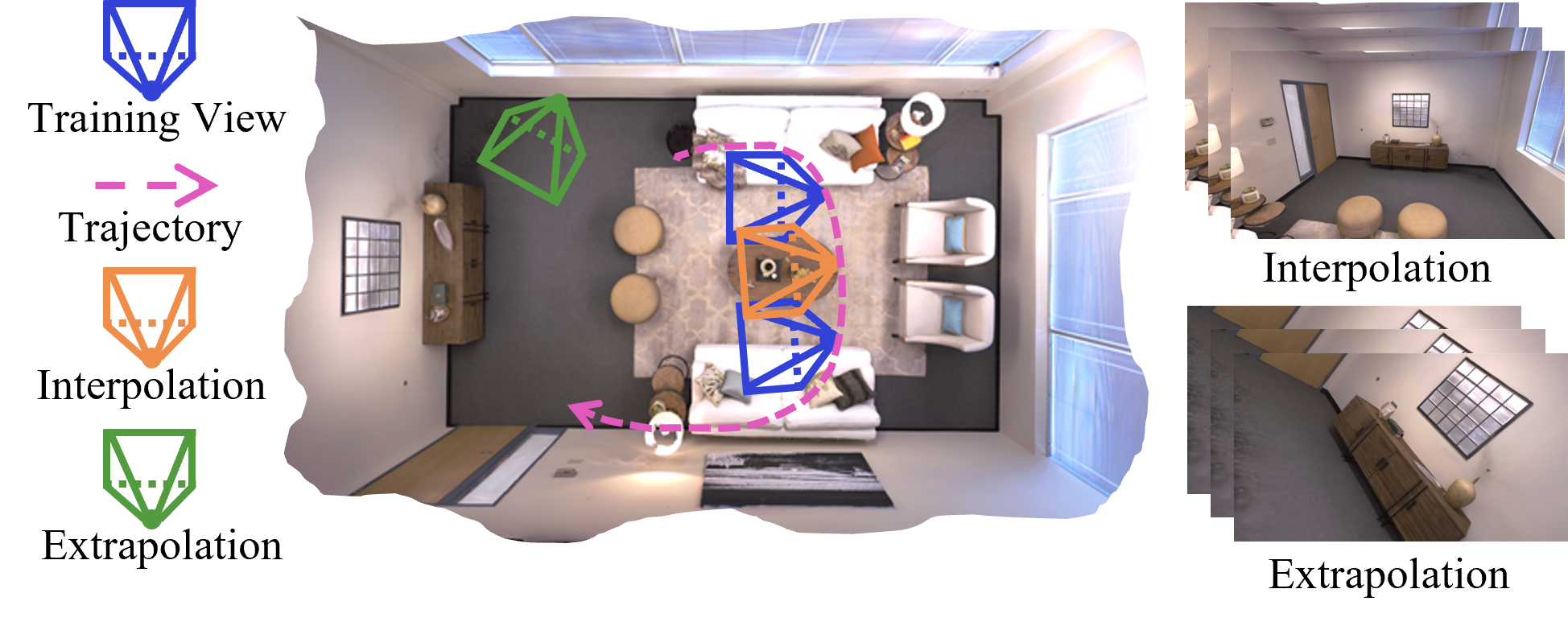}   
	\caption{Illustration of view interpolation and extrapolation. Interpolated views closely resemble the training views, whereas extrapolated views differ significantly from them. 
	\label{fig:inter-extra}} 
    \vspace{-1.7em}
\end{figure}

Albeit inspiring, many viewpoints inevitably remain unobserved even with crowd-sourcing. Exhaustively covering every possible view would be prohibitively expensive in terms of time and resources. This introduces a critical challenge: maintaining high visual quality from multiple viewpoints, many of which have never been directly observed by contributors -- a problem known as novel view synthesis (NVS). 
NVS can be further categorized into two groups based on the similarity between the input view and the novel view: interpolation and extrapolation~\cite{nerfvs, vegs}. Interpolation focuses on synthesizing novel views that are similar to the training views collected from contributors, while extrapolation addresses the synthesis of views that are significantly different from training views, as shown in Fig.~\ref{fig:inter-extra}, and is more challenging.  
In map-sharing systems, extrapolation is particularly common, as contributors often only partially cover the scene. 
A few studies have sought to address this issue: some focus on novel view synthesis with few-shot learning~\cite{loopsparsegs, fsgs, fewshotDepthRegularized, dngaussian, splatt3r}, while others explore view generalization across different scenarios~\cite{pixelsplat, transplat, mvsplat, latentsplat, adgaussian}. Although these works improve general NVS performance,  
they still suffer from limited rendering quality and can only handle a small number of input images.
In parallel, some prior works have investigated view extrapolation using implicit representations~\cite{nerfvs,ying2023parf},  but these methods are not applicable to Gaussian splatting due to their different representations. 
As a result, a high-fidelity Gaussian-splatting-based map-sharing system remains elusive.

In addition to the need for view extrapolation, a practical map-sharing system has another important feature: it must deliver maps to users as efficiently as possible. 
Given that 3D Gaussian maps are typically large -- e.g., 0.3~GB for a small single room in the Replica dataset~\cite{replica} -- transmitting raw Gaussians can consume significant traffic and reduce user experiences.
The situation becomes even more challenging when the system must incorporate progressively collected observations from contributors. 
Such a system must continuously update the global map and efficiently distribute these updates to users with minimal overhead. 
To address the transmission bottleneck, several approaches have been developed, including anchor-based methods~\cite{scaffold-gs, hac, octree, hac++},  pruning-based methods~\cite{lightgaussian, radsplat, minisplatting, lp3dgs}, and quantization-based  methods~\cite{compressionReview, hac, compact3d1, compact3d2, lightgaussian, EAGLES, RDOGaussian}. 
However, these methods are primarily designed for a single, static map, the challenge of continuously updating the global map and distributing it to users remains largely unexplored.

\textit{Given the above challenges -- specifically, how can we design a practical framework to construct a high-fidelity Gaussian map collected from contributors, and efficiently share the continuously updating map with users?}
In this paper, we address this problem by proposing \mysystem. 
Following the server-client architecture used in Map++~\cite{map++}, \mysystem maintains a global map on the server, with each user retaining a local copy. Users can volunteer to become contributors. As contributors explore their surroundings, they upload images from various perspectives to the server, enabling the global map to be enhanced and shared continuously.     
To facilitate such a pipeline, \mysystem employs a modular design that accurately registers both contributors and users to the system, preserves a compact yet expressive representation of the global map, and enhances map quality through customized training strategies. The proposed pipeline enables high-quality map sharing and efficient distribution to users, fulfilling the core goals of \mysystem. 
To further enhance the view extrapolation capability of the global map, \mysystem constructs an auxiliary virtual map from the input data, from which various virtual images are sampled. Each virtual image is associated with predicted per-pixel confidence scores. Together, the generated virtual images and confidence scores serve as pseudo ground truth, augmenting the training data and ultimately improving map quality.  
Once the global map is constructed, \mysystem supports on-demand map access: users upload their current view, which the server matches against contributor images to localize the user. Once matched, the corresponding map segment is transmitted. For the initial transmission, a full map will be delivered. 
For continuous map updates, \mysystem employs a lightweight representation named a map increment. Instead of retraining and transmitting the entire global map to replace the outdated map, \mysystem maintains a database of staged maps and updates the map database by training and transmitting only the incremental changes. 
In this way, the cached map on user devices can be effectively reused. By integrating the received increment with the cached data, users can reconstruct the up-to-date global map without performance degradation.

We summarize our contributions as follows:
\begin{itemize}
    \item  
    To the best of our knowledge, \mysystem is the first Gaussian map-sharing framework that supports both continuous map updates and novel view synthesis. As more contributor data is integrated, the quality of the shared map progressively improves with minimal transmission overhead.
 
    \item To enable high-fidelity and compact map sharing, we develop a complete and modular pipeline with customized representations and training strategies. In addition to the pipeline, we also incorporate a set of key techniques, including virtual-image-based map enhancement for improved novel view synthesis and incremental map updates for low-overhead synchronization between the global map and user-side local maps.
 
    \item We evaluate \mysystem on a benchmark built from the Replica dataset~\cite{replica}, and the results show that, compared to the state-of-the-art method, \mysystem improves PSNR, LPIPS, and Depth L1 by an average of 11\%, 22\%, and 74\% on extrapolated views, respectively, while also being more compact, reducing map transmission overhead by 36\%.
\end{itemize}

\section{Related Work}\label{sec:related}
\subsection{Map Sharing}
 
Several recent studies have explored map sharing among multiple users. 
Map++~\cite{map++} adopts a server-client-based framework, where a global map is constructed on the server, and each user maintains a partial copy of the global map. 
SLAM-Share~\cite{slamshare} offloads most of the SLAM computation to the server, enabling users to efficiently leverage the shared map. 
Pair-Navi~\cite{pairnavi1, pairnavi2} constructs a trajectory map on the server, and assists subsequent users by reusing the experience of previous travelers.
RecNet~\cite{recnet} represents map data as range images and latent vectors for efficient transmission.
Google ARCore~\cite{googleAR} maintains a sparse anchor map on the server, enabling localization through anchor sharing.
These methods primarily focus on map sharing for localization or sparse reconstruction, while \mysystem is designed for dense reconstruction with 3D Gaussians.   

\subsection{Novel View Synthesis} 
Several recent efforts have been made in novel view synthesis. For instance, LoopSparseGS~\cite{loopsparsegs} and DNGaussian~\cite{dngaussian} address the challenge of sparse observations by introducing a depth-aware regularization loss and leveraging geometric cues.
Some works focus on view generalization cross scenarios~\cite{pixelsplat, transplat, mvsplat, latentsplat, adgaussian, anysplat}. For example, 
PixelSplat~\cite{pixelsplat} estimates the parameters of a 3D Gaussian for each scenario using a single image pair in a single forward pass, while  
MVSplat~\cite{mvsplat} adopts a similar approach but extends it to multi-view images.
Although these works improve general novel view synthesis performance, they still suffer from limited rendering quality and can only handle a small number of input images. 
For view extrapolation, NerfVS~\cite{nerfvs} proposes leveraging depth priors and view coverage priors to guide optimization in NeRF, while PARF~\cite{ying2023parf} fuses semantic, primitive, and radiance information into a single framework, enabling high-fidelity and high-speed rendering. Although effective, these methods are primarily designed for radiance fields, which are not directly applicable to Gaussian splatting.

\subsection{Compact 3D Gaussian Splatting}
To compress the Gaussian map, several approaches have been developed, which can be classified into three categories: 
(1) Anchor-based methods~\cite{scaffold-gs, hac, hac++, octree, instancegaussian, fcgs, pcgs}, which encodes high-dimensional Gaussian attributes, such as spherical harmonics, positions, colors, and opacities into low-dimensional anchor features; 
(2) Pruning~\cite{hac, lightgaussian, radsplat, minisplatting, lp3dgs}, which filters out less critical Gaussians based on importance scores.  
and (3) Quantization~\cite{compressionReview, hac, compact3d1, compact3d2, lightgaussian, EAGLES, RDOGaussian, fcgs}, which maps high-precision Gaussian parameters to lower-precision discrete levels to reduce storage and bandwidth.
Among these works, HAC~\cite{hac} is most closely related to ours, it incorporates all three techniques within a unified framework, guided by a context model that dynamically adjusts quantization step sizes.  
Although these methods effectively reduce the size of 3D Gaussians, they are primarily designed for single-user scenarios with static maps. A suitable representation for continuously updated, multi-user map-sharing systems remains an open challenge.
 
\begin{figure*}[t]
	\centering
	\includegraphics[width = 0.98\linewidth]{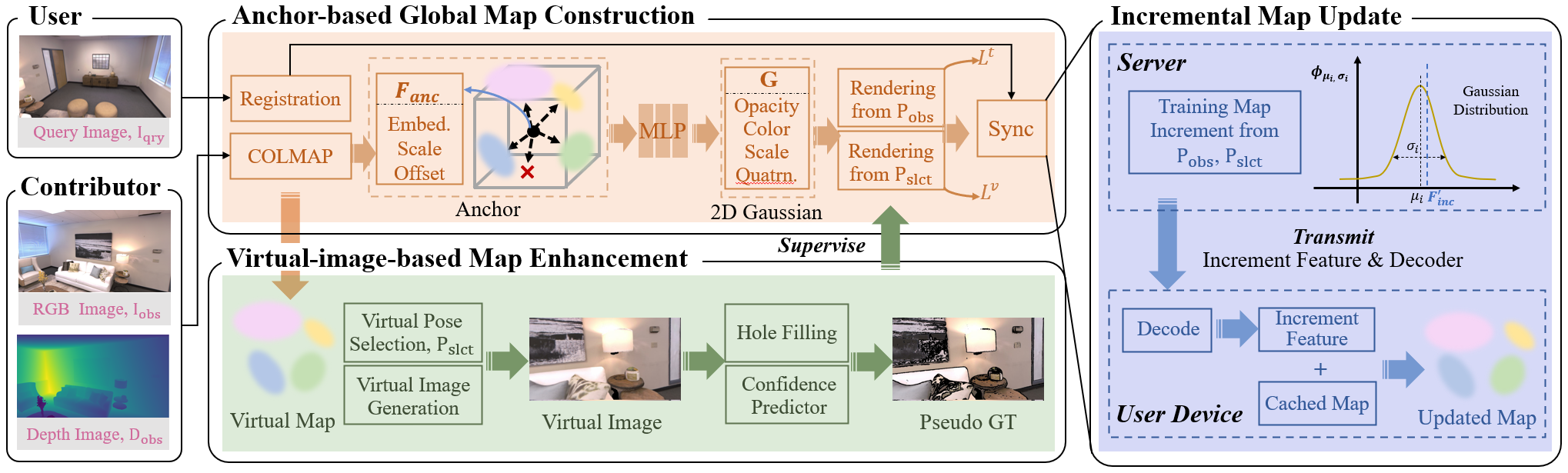}
	\caption{Overview of \mysystem.
        \mysystem takes RGB-D images as input and aligns them to a unified coordinate system using COLMAP~\cite{colmap}, followed by a global map construction process. 
        Since the global map must be transmitted to users, its size is critical, motivating a compact representation. 
        To further enhance map quality, \mysystem generates an auxiliary virtual map and renders virtual images from it.  
        After post-processing, these virtual images serve as pseudo ground truth to augment the training data. Once the global map is reconstructed, users register with the server to access it. \mysystem transmits the full map during the initial stage and the map increments in subsequent updates. By leveraging the received increments, users reconstruct the latest Gaussian map with minimal transmission overhead.  \label{fig:overview}}
   \vspace{-1.7em}
\end{figure*}

\section{Preliminary on Anchor-based 3D Gaussian Splatting}\label{sec:preliminary}
Recently, 3D Gaussian Splatting (3DGS)~\cite{3DGS} has gained considerable attention as an effective solution for 3D reconstruction. It models the scene using a collection of anisotropic 3D Gaussian ellipsoids, enabling high-quality rendering.  
A 3D Gaussian ellipsoid is formulated as: 
    \begin{equation}
    G(x) = e^{-\frac{1}{2}(x-\mu)^T\Sigma^{-1}(x-\mu)}, \Sigma = RSS^TR^T, 
    \end{equation} 
where $x \in \mathbb{R}^3$ is a spatial point, $\mu \in \mathbb{R}^3$ is the Gaussian center, and $R$, $S$ denote rotation and scaling, respectively. 
Gaussians are then rendered onto the image plane via $\alpha$-blending:
\begin{equation}\label{equ:3DGSrendering} 
    C=\sum_{i \in I}c_i\sigma_i\prod \limits_{j=1}^{i-1}(1-\sigma_j),   
\end{equation} 
where $C$ is the rendered pixel color along a ray, $I$ is the set of Gaussians that intersect the ray. $c_i$ and $\sigma_i$ denote the color and opacity of the $i$-th Gaussian, respectively.
To further compress 3DGS, an anchor-based representation~\cite{hac} encodes color, opacity, and rotation into a low-dimensional feature embedding $F_{emb}$. 
Assuming each anchor is responsible for $K$ nearby Gaussians, it can be defined as follows: 
\begin{equation} \label{equ:anchorFeature} 
    F_{anc} = \{ F_{s},F_{o}, F_{emb}\},   
\end{equation}
where $F_{s} \in R^{K \times 3}$ and $F_{o} \in R^{K \times 3}$ represent the scales and offsets of $K$ corresponding Gaussians, respectively, and $F_{emb} \in R^{50}$ denotes the feature embedding. The compressed anchor $F_{anc}$ can be decoded back into 3D Gaussians:  
\begin{equation} \label{equ:no-incremental}  
   G = Decode(F_{anc}) , 
\end{equation}
where $G$ represents the recovered Gaussians. A two-layer MLP is employed to decode $F_{emb}$ back into Gaussian attributes, and $F_{o}$ is added to the anchor position $V$ to restore Gaussians' 3D positions.
 
\section{Method}
\label{sec:method}
This section details the design of \mysystem. As shown in Fig.~\ref{fig:overview}, \mysystem takes observations from contributors as input and constructs a global map and a virtual map. 
The design goals of \mysystem are two-fold: first, the constructed map should be of high accuracy; Second, it should be continuously updated and efficiently distributed to users. To achieve these goals, \mysystem incorporates three modules: anchor-based global map construction, virtual-image-based map enhancement, and incremental map update. 
 
\subsection{Anchor-based Global Map Construction} \label{subsec:anchor}  
We first present a comprehensive pipeline for global map construction, designed to support high-fidelity, compact map sharing across multiple contributors.
To initiate this process, \mysystem collects multi-view images from contributors who explore the environment and upload their observations to the server.
To align data from different contributors into a unified global coordinate system, \mysystem employs COLMAP~\cite{colmap}, a Structure-from-Motion (SfM) module,  to estimate camera poses from the collected images. 
Based on these poses, a compact global map is constructed using an anchor-based Gaussian representation. 
Specifically, \mysystem first lifts all the input images into 3D space and generates colored point clouds, which are then voxelized as follows:  
\begin{equation}  
    V = \{|\frac{P_{raw}}{\epsilon}|\} \cdot \epsilon,  
\end{equation}
where $V$ represents the anchor positions, $P_{raw}$ is the point cloud coordinate, and $\epsilon$ is the voxel resolution (e.g., 3cm in our case). 
These anchors are then used to initialize the global map following HAC~\cite{hac}. 
However, naive 3D Gaussian splatting exhibits significant multi-view inconsistency~\cite{2DGS}: rendering a 3D Gaussian ellipsoid from different perspectives yields different results. We observed that such inconsistency can severely degrade map quality, especially for view extrapolation. 
To resolve this, we replace 3D ellipsoids with view-consistent 2D flats and incorporate a normal loss, following~\cite{2DGS}.
Experiments show that this approach improves map quality and reduces overall map size.  
While beneficial, we further observe that this strategy alone is insufficient: inaccurate normal estimates near object boundaries can still degrade view extrapolation, particularly when input data is insufficient. To mitigate this, \mysystem identifies regions with a depth L1 error exceeding a threshold (e.g., 0.1m) as unreliable and excludes them from the normal loss computation, thereby further improving the performance. 
Finally, the loss function for the training/input views is defined as follows:
\begin{equation}   \label{eq:trainingloss} 
    L^t = L1_{obs}^t +  L_{SSIM}^t +  L_{reg}^t +  L_d^t +   L_{nm}^t + L_{mask}^t,  
\end{equation}
where $L1_{obs}^t$ and $L_{SSIM}^t$ denote the L1 and SSIM loss~\cite{ssim} between the rendered image and the ground truth image, respectively. $L_{reg}^t$ denotes the regularization term for scale~\cite{volume}, $L_d^t$ is the L1 loss between the rendered depth and ground truth depth, $L_{nm}^t$ accounts for the normal loss~\cite{2DGS} after edge filtering, 
$L_{mask}^t$ refers to the learning-based Gaussian pruning loss~\cite{compact3d1}. Please note that each loss term is associated with a weight, the specifics of which are provided in the implementation section.
 
Once training completes, the global map becomes available for use. During inference, \mysystem performs image-based registration~\cite{opencv} by matching a user’s uploaded view against contributor images with known poses. Based on the best match, the system estimates the user's location and transmits the corresponding map segment.   
Altogether, \mysystem offers a unified framework for constructing and sharing 3D Gaussian maps, laying the foundation for subsequent refinement and incremental updates.

\begin{figure}[t]
	\centering
	\includegraphics[width = \linewidth]{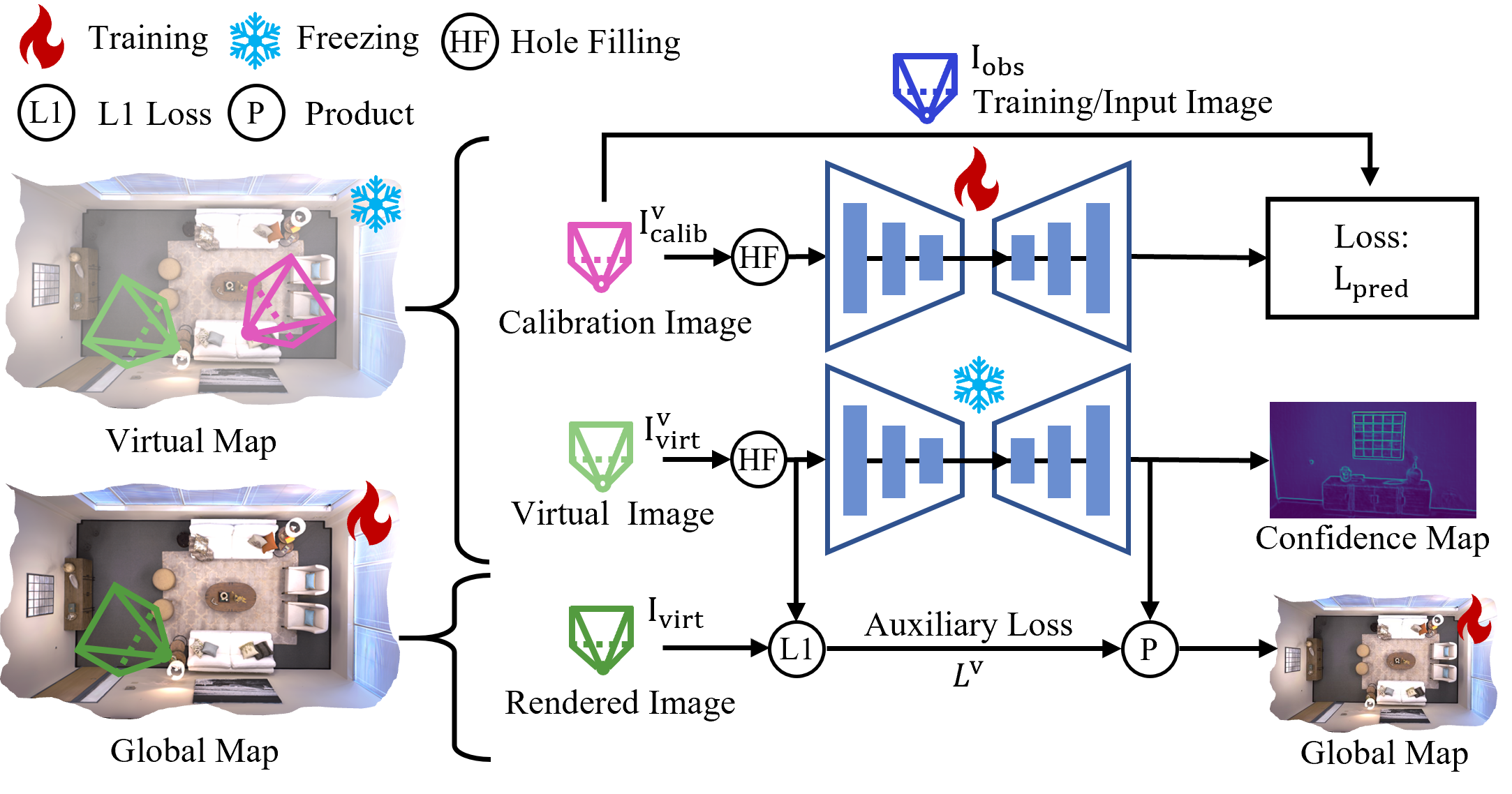}
	\caption{The process of virtual-image-based map enhancement. It uses $L_{pred}$ as the loss for training the predictor and $L^{v}$ for training the global map. During the training of the global map, the parameters of the predictor are frozen. }
	\label{fig:train}  
      \vspace{-1.7em}
\end{figure}

\subsection{Virtual-image-based Map Enhancement}\label{subsec:virtual}  
As map-sharing systems demand high reusability, they place stricter requirements on novel view synthesis. To further enhance the quality of the shared map, we introduce an auxiliary virtual map to augment the global map. Similar to the global map, the virtual map is constructed from the input observations and can be rendered into images.  These rendered images act as pseudo ground truth to guide the training of the global map.  
Although both maps are derived from the same input, the virtual map is solely responsible for enhancing the global map, allowing it to be explicitly optimized for preserving fine-grained geometric cues from the raw sensor data. In contrast, the global map -- trained using the standard 3DGS pipeline -- tends to lose such detail due to overfitting to the input views. 
To construct such virtual map, we first lift the RGB-D images into 3D space to generate colored point clouds. 
These point clouds are then converted into 3D Gaussians. 
The attributes of each Gaussian are computed as follows:  the position and color of each Gaussian are directly derived from the point cloud; the scale is determined based on the point cloud density~\cite{3DGS}; the opacity is set to 1; and all Gaussians are isotropic, with rotation not considered. We intentionally avoid using 2D Gaussians here, as they require extensive training data for reliable rotation estimation, whereas a properly initialized isotropic 3DGS can generalize well to novel views even without training. The resulting virtual map can then be rendered.

Since interpolated views closely resemble the input images and require little augmentation, we render virtual images only for extrapolated views to serve as pseudo ground truth.  
Specifically, after obtaining camera poses via COLMAP~\cite{colmap}, virtual poses are randomly sampled across the entire scene. We retain only the poses that qualify as extrapolated views: those that are sufficiently different from all input poses are kept, while any pose that is too close (e.g., with a rotation difference less than 10 degrees) is discarded.  
Based on the selected poses $P_{slct}$, virtual images $I^v_{virt}$ are rendered, where the superscript $v$ indicates that the image is rendered from the virtual map. 
The rendered virtual images $I^v_{virt}$, however, require further post-processing to serve as reliable pseudo ground truth.   We devise two techniques to improve the quality of $I^v_{virt}$: hole filling and confidence prediction. 
We observe that when input data is insufficient, certain regions are not properly covered by Gaussians, resulting in visible holes. To address this,  
we render an additional opacity image alongside $I^v_{virt}$ to identify those poorly covered areas.
Regions where the opacity falls below a given threshold are marked as holes and filled using the Navier-Stokes inpainting algorithm~\cite{navier}.  
Even after hole filling, the virtual image may still be inaccurate. To account for this, we incorporate a confidence prediction step to assess the reliability of each pixel and guide its use in supervision. 
Those areas with a low confidence score will be assigned a smaller weight when supervising the global map.  
To train the confidence predictor, we render calibration images $I^v_{calib}$ from the virtual map at the training view (input pose $P_{obs}$), where accurate ground truth is available for supervision.
The predictor takes $I^v_{calib}$ as input and predicts a dense per-pixel confidence map. The loss function is defined as follows:  
\begin{equation} \label{eq:predloss}
    L_{pred} = \left| \text{Pred}(I^v_{calib}) - \frac{\tau - \text{L1}^v_{calib}}{\tau} \right|,
\end{equation}
where $ L_{pred}$ is the loss for training the predictor, $\text{Pred}(I^v_{calib})$ is the estimated confidence on calibration image $I^v_{calib}$,  $\text{L1}^v_{calib}$ denotes the accuracy of $I^v_{calib}$, which is the L1 error between the calibration image $I^v_{calib}$ and the input image $I_{obs}$, and $\tau$ is the maximum $\text{L1}^v_{calib}$ of all calibration images in the dataset. As shown in Fig.~\ref{fig:train}, the predictor is built upon U-Net~\cite{unet}, consisting of an encoder-decoder structure with skip connections to preserve spatial details.  

Once trained on calibration images from the training views, the predictor is applied to virtual images $I^v_{virt}$ rendered from extrapolated viewpoints. 
The global map can be then enhanced using the following auxiliary loss:
\begin{equation}  \label{eq:virtualloss}
    L^{v} =  \text{L1}^v_{virt}  \cdot   \text{Pred}(I^v_{virt}) , 
\end{equation}
where $\text{L1}^v_{virt}$ denotes the L1 error between the virtual image $I^v_{virt}$ and the corresponding rendered image $I_{virt}$ from the global map, and $\text{Pred}(I^v_{virt})$ is the predicted confidence map for $I^v_{virt}$. 
This confidence map encourages the global map to focus on high-confidence regions.
The total loss for global map construction is thus:
\begin{equation} \label{eq:total_loss}
    L^{total} = L^t + L^v,
\end{equation}
where $L^t$ is the loss for input views, detailed in Eq.~\eqref{eq:trainingloss} and $L^v$ is the auxiliary loss for virtual views. The associated loss weights will be provided in the implementation section.
 
\subsection{Incremental Map Update} \label{subsec:incremental} 
\begin{figure}[t]
	\centering
	\includegraphics[width = \linewidth]{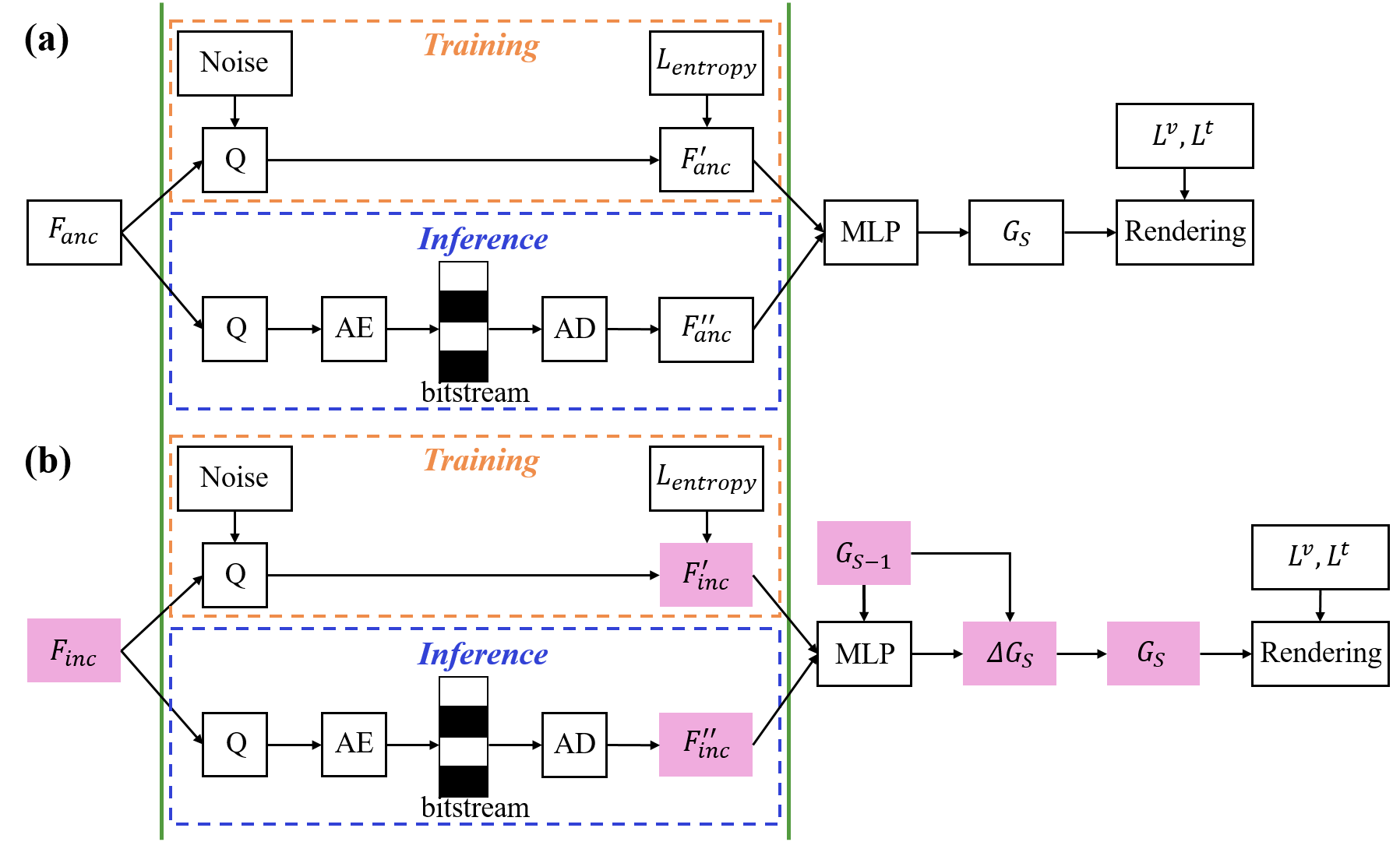}
	\caption{The process of training and transmitting map data. (a) and (b) represent the procedures without and with incremental map update, respectively. AE refers to arithmetic coding~\cite{arithmetic}, while AD denotes the corresponding decoding procedure. }
	\label{fig:quantization}
    \vspace{-1.7em}
\end{figure}

Next, we describe how the shared map is efficiently transmitted to the user.
Following Map++~\cite{map++}, \mysystem categorizes the global map into seen and unseen areas. 
For unseen areas, a full map segment is initially transmitted. 
For seen areas, the map is progressively refined as the system progresses. 
To support lightweight synchronization and efficient updates, \mysystem proposes to update the global map in the form of map increments.
A map increment ${F_{Inc}}$ consists of the same components as $F_{anc}$ in Eq.~\eqref{equ:anchorFeature}, but is explicitly designed to be more compact: the feature embedding dimension in ${F_{Inc}}$ is reduced by half compared to $F_{anc}$.  Moreover, the process of translating map data into 3D Gaussians is also different, as illustrated in Fig.~\ref{fig:quantization}. 
Specifically, ${F_{Inc}}$ is first quantized into $ {F_{Inc}}'$, which is then decoded into a Gaussian increment $\Delta G_S$:     
\begin{equation} \label{equ:incremental2}
\begin{split} 
    \Delta G_S &= Decode({F_{Inc}}'), \\ 
\end{split}  
\end{equation}  
where ${F_{Inc}}'$ is the quantized map increment, $S$ denotes the stage ID of the map update, which is initialized to 0 and incremented by one for each stage. 
In \mysystem, each stage advances by incorporating additional observations, specifically those collected from a single contributor. 
$Decode$ denotes the decoding operation, which follows the same design as described in Sec.~\ref{sec:preliminary}. 
Finally, the decoded Gaussian increment $\Delta G_S$ is integrated with the outdated map $G_{S\text{-}1}$ from the previous stage to derive the updated map $G_S$. The process is formulated as follows: 
\begin{equation} 
    G_{S} =  G_{S-1} + \Delta G_S.
\end{equation}  
Here, the addition denotes element-wise operations across the corresponding attributes of each Gaussian.

Next, we describe the training procedure for the map increment ${F_{Inc}}$, as illustrated in Fig.~\ref{fig:quantization}.  To enable adaptive quantization, ${F_{Inc}}$ is initialized as an empty embedding, along with a learnable quantization step ${st}$~\cite{hac}. 
To allow differentiable training of $st$, we follow the strategy proposed in~\cite{variational}, adding noise during training and applying rounding only at inference time. 
Specifically, we inject uniform noise sampled from the interval  $[\text{-}\frac{1}{2}{st}, \text{+}\frac{1}{2}{st}]$ into each element of ${F_{Inc}}$, simulating the effect of quantization and get ${F_{Inc}}'$ as an intermediate result. The quantized feature ${F_{Inc}}'$ is then decoded into Gaussian maps and supervised using the following rate-distortion loss:    
\begin{equation} 
    L_{updt} = \lambda_{q}  L_{entropy} +  L^{total},  
\end{equation} 
where $L_{entropy}$ is the bit rate estimated using the entropy model following HAC~\cite{hac}, $\lambda_{q}$ is the corresponding loss weight, and $L^{total}$ denotes the distortion term defined in Eq.~\eqref{eq:total_loss}.
After training, we obtain both the optimized quantization step $st$ and the trained map increment ${F_{Inc}}$.
To efficiently transmit ${F_{Inc}}$ to users, we perform a series of post-training steps, collectively referred to as the inference stage. Unlike the   noise injection used during training, quantization in the inference stage is performed by directly rounding each element of ${F_{Inc}}$ as follows: 
\begin{equation}\label{equ:quantization}
    {F^{\sigma}_{Inc} = \{|\frac{{F_{Inc}}}{{st}}|\} \cdot {st}},  
\end{equation} 
where $F^{\sigma}_{Inc}$ denotes the final quantized map increment. This result is further compressed into a bitstream using arithmetic coding (AE)~\cite{arithmetic}. The compressed bitstream, along with the map increment decoder, is then transmitted to users and translated into the updated Gaussian map.   
\section{Experimental Evaluation}
\label{sec:experiment}  
We first describe our experimental setup.
We compare \mysystem against two baselines: the default 3DGS~\cite{3DGS} and  HAC~\cite{hac}, a state-of-the-art compression method.
Other methods evaluated in HAC~\cite{hac} are not included, as their performance has already been thoroughly analyzed. 
To ensure a fair comparison, we implement a server-client-based map-sharing framework following Map++~\cite{map++}, and port both 3DGS and HAC into this framework as baselines.

\subsection{Implementation} We implement \mysystem in PyTorch~\cite{pytorch} and train it on a single NVIDIA RTX 3090Ti GPU. The voxel size $\epsilon$ is set to 0.03m, with each anchor representing 10 Gaussians.   
We adopt the SfM-based COLMAP~\cite{colmap} for pose estimation. Compared to the SLAM-based method~\cite{yugay2023gaussian}, COLMAP yields more accurate poses due to its longer optimization time.
For each stage, the predictor is trained for 3,000 iterations (approximately 15 minutes), and the global map is trained for 12,000 iterations (around 30 minutes).
Training stops once the map quality, measured by PSNR, converges.
The predictor follows an encoder-decoder architecture, downsampling the image by a factor of 8 before restoring it to the original resolution.
During the training of the global map, we set the loss weights to: $\lambda_{obs}^t=1$, $\lambda_{SSIM}^t=0.05$, $\lambda_{reg}^t=0.1$, $\lambda_{d}^t=1.0$, $\lambda_{nm}^t=0.1$, $\lambda_{mask}^t=0.005$, $\lambda^t=1.0$, $\lambda^v=0.1$, and $\lambda_{q}=0.0025$. 

\begin{table*}[h]
\centering 
\setlength\tabcolsep{5.0pt} 
\renewcommand\arraystretch{1} 
\caption{\label{tab:overall_extra} Quantitative results. We evaluated the performance of HAC~\cite{hac}, 3DGS~\cite{3DGS}, and \mysystem across 8 scenes in \mydataset on extrapolated views. The best results are highlighted in {\colorbox{best}{red}}.} 
\vspace{-4pt}
\begin{tabular}{c||l|ccccccccc} 
\hline \hline
\textbf{Method} & \textbf{Metric} & \textbf{Office0} & \textbf{Office1} & \textbf{Office2} & \textbf{Office3} & \textbf{Office4} & \textbf{Room0} & \textbf{Room1} & \textbf{Room2} & \textbf{Average}  \\
\hline
\hline
\multirow{5}{*}{\makecell[c]{\textbf{HAC~\cite{hac}} \\ \textit{extrapolation}}}& PSNR [dB]$\uparrow$ & 31.91  & 33.43  &27.36 & 26.82  & 28.95   & 26.10 &27.64 & 25.79 & 28.50 \\
                        & LPIPS$\downarrow$         & 0.121 & 0.117 & 0.161 &  0.134   &  0.148  & 0.184 & 0.179    &0.159 & 0.150 \\
                        & SSIM $\uparrow$            & 0.953 & 0.970 & 0.929 & 0.941   &0.953   &0.898 & 0.918     &0.924  & 0.936\\
                        & Depth L1 [cm]$\downarrow$         & 1.76 & 2.25 & 3.66 & 5.98  & 2.99 &  2.80 & 1.82  & 9.87 & 3.89\\
                        & Size [MB]$\downarrow$         & 3.15   & 1.47  & 4.33  & 4.23 & 3.98 & 5.81 & 4.54  & 2.64 & 3.76\\
\hdashline

\multirow{5}{*}{\makecell[c]{\textbf{3DGS~\cite{3DGS}} \\ \textit{extrapolation}}}& PSNR [dB]$\uparrow$ & 31.46  & 33.58  & 27.86 & 27.06 & 29.18 & 26.13 & 27.67 & 27.21 & 28.77 \\
                        & LPIPS $\downarrow$        & \cellcolor{best}0.093  & \cellcolor{best}0.088  & 0.132 & 0.115 & 0.120   & 0.164 & \cellcolor{best}0.154 & 0.124 & 0.124 \\
                        & SSIM $\uparrow$           & 0.948  & 0.970  & 0.926 & 0.935 & 0.949   & 0.897 &0.911 & 0.935 & 0.934\\
                        & Depth L1 [cm]$\downarrow$        & 1.50  & 1.82  & 2.68 & 6.54 & 5.00   & 2.37 & 1.57 & 11.53 & 4.13\\
                        & Size [MB] $\downarrow$        & 184.99  & 93.60  & 238.59 & 261.26 & 291.83   & 310.85 & 242.61 & 140.12 & 220.48\\
\hdashline

\multirow{5}{*}{\makecell[c]{\textbf{\mysystem} \\ \textit{extrapolation}}}& PSNR [dB]$\uparrow$& \cellcolor{best}33.66 & \cellcolor{best}34.90 & \cellcolor{best}31.60 & \cellcolor{best}30.53 &  \cellcolor{best}32.95  & \cellcolor{best}28.83 & \cellcolor{best}29.18 & \cellcolor{best}30.69 & \cellcolor{best}31.54 \\
                        & LPIPS$\downarrow$      & 0.103 & 0.107& \cellcolor{best}0.102  & \cellcolor{best}0.092 & \cellcolor{best}0.110& \cellcolor{best}0.145  &0.159   &  \cellcolor{best}0.113     & \cellcolor{best}0.117 \\
                        & SSIM$\uparrow$      & \cellcolor{best}0.965 & \cellcolor{best}0.978& \cellcolor{best}0.975  & \cellcolor{best}0.968 & \cellcolor{best}0.974& \cellcolor{best}0.937  &\cellcolor{best}0.929   & \cellcolor{best}0.966      & \cellcolor{best}0.962\\
                        & Depth L1 [cm]$\downarrow$        & \cellcolor{best} 0.76 & \cellcolor{best}0.65& \cellcolor{best}0.51  & \cellcolor{best}1.60 & \cellcolor{best}1.14 & \cellcolor{best}1.08  &\cellcolor{best}0.85  & \cellcolor{best}1.40       & \cellcolor{best}1.00\\
                        & Size [MB]$\downarrow$    & \cellcolor{best}1.93   & \cellcolor{best}0.94   & \cellcolor{best}2.78  & \cellcolor{best}2.60  & \cellcolor{best}2.47   & \cellcolor{best}3.72   & \cellcolor{best}2.86   & \cellcolor{best}2.04    & \cellcolor{best}2.42\\

\hline \hline 

\end{tabular}
\end{table*}  

\begin{table*}[h]
\centering 
\setlength\tabcolsep{5.0pt} 
\renewcommand\arraystretch{1} 
\caption{\label{tab:overall_inter} Quantitative results. We evaluated the performance of HAC~\cite{hac}, 3DGS~\cite{3DGS}, and \mysystem across 8 scenes in \mydataset on interpolated views. The best and second-best results are highlighted in {\colorbox{best}{red}} and {\colorbox{bestt}{orange}}, respectively. } 
\vspace{-4pt}
\begin{tabular}{c||l|ccccccccc} 
\hline \hline
\textbf{Method} & \textbf{Metric} & \textbf{Office0} & \textbf{Office1} & \textbf{Office2} & \textbf{Office3} & \textbf{Office4} & \textbf{Room0} & \textbf{Room1} & \textbf{Room2} & \textbf{Average}  \\
\hline
\hline

\multirow{5}{*}{\makecell[c]{\textbf{HAC~\cite{hac}} \\ \textit{interpolation}}}& PSNR [dB]$\uparrow$ & \cellcolor{bestt}38.63   &  \cellcolor{bestt}39.92 & 34.49 & 34.77 &  35.19  & \cellcolor{bestt}31.62 &\cellcolor{bestt}35.23 &35.34 & 35.65 \\
                        & LPIPS$\downarrow$        &  0.091   & 0.083 & 0.123 &0.070    & 0.101 & 0.094 & 0.112 &  0.089 & 0.095 \\
                        & SSIM$\uparrow$          & \cellcolor{bestt}0.987    & \cellcolor{bestt}0.994 & 0.984 & 0.990 &  0.985 & 0.978 & \cellcolor{bestt}0.982 & 0.989 & 0.986\\
                        & Depth L1 [cm]$\downarrow$        &  \cellcolor{bestt}0.48  & \cellcolor{bestt}0.31 & 0.46 &  \cellcolor{bestt}0.88  & 0.79 & \cellcolor{bestt}0.76  & 0.43 & \cellcolor{bestt}0.97 & 0.64\\
\hdashline

\multirow{5}{*}{\makecell[c]{\textbf{3DGS~\cite{3DGS}} \\ \textit{interpolation}}}& PSNR [dB]$\uparrow$ & \cellcolor{best}40.76  & \cellcolor{best}42.01  & \cellcolor{best}36.14 & \cellcolor{best}36.67 & \cellcolor{best}37.48   & \cellcolor{best}33.33 & \cellcolor{best}37.44 & \cellcolor{best}37.33 & \cellcolor{best}37.64 \\
                        & LPIPS $\downarrow$        & \cellcolor{best}0.036  & \cellcolor{best}0.028  & \cellcolor{best}0.041 & \cellcolor{best}0.026 & \cellcolor{best}0.042   & \cellcolor{best}0.044 & \cellcolor{best}0.048 & \cellcolor{best}0.036 & \cellcolor{best}0.038 \\
                        & SSIM $\uparrow$           & \cellcolor{best}0.993  & \cellcolor{best}0.996  & \cellcolor{best}0.992 & \cellcolor{best}0.995 & \cellcolor{best}0.992   & \cellcolor{best}0.985 &\cellcolor{best}0.990 & \cellcolor{best}0.994 & \cellcolor{best}0.992\\
                        & Depth L1 [cm]$\downarrow$        & \cellcolor{best}0.35  & \cellcolor{best}0.22  & \cellcolor{best}0.28 & \cellcolor{best}0.52 & \cellcolor{best}0.49   & \cellcolor{best}0.62 &\cellcolor{best}0.25 & \cellcolor{best}0.68 & \cellcolor{best}0.43\\

\hdashline
                        
\multirow{5}{*}{\makecell[c]{\textbf{\mysystem} \\ \textit{interpolation}}}& PSNR [dB]$\uparrow$  & 38.54 & 39.73 & \cellcolor{bestt}34.70     &  \cellcolor{bestt}34.98 & \cellcolor{bestt}35.97 & 31.60 & 34.96 & \cellcolor{bestt}35.77 & \cellcolor{bestt}35.78 \\
                          & LPIPS$\downarrow$      & \cellcolor{bestt}0.082 & \cellcolor{bestt}0.078 & \cellcolor{bestt}0.095    & \cellcolor{bestt}0.057 & \cellcolor{bestt}0.081 & \cellcolor{bestt}0.082 & \cellcolor{bestt}0.105 & \cellcolor{bestt}0.075 & \cellcolor{bestt}0.082 \\
                          & SSIM$\uparrow$      & 0.986 & 0.992 & \cellcolor{bestt}0.987    & \cellcolor{bestt}0.992 & \cellcolor{bestt}0.989 & \cellcolor{bestt}0.979 & 0.981 & \cellcolor{bestt}0.990 & \cellcolor{bestt}0.987\\
                          & Depth L1 [cm]$\downarrow$        & 0.49 & 0.33 & \cellcolor{bestt}0.42    & 0.93 &  \cellcolor{bestt}0.68 & 0.77 &  \cellcolor{bestt}0.37 &  1.06 &  \cellcolor{bestt}0.63\\

\hline \hline 

\end{tabular}
\vspace{-1em}
\end{table*}

\subsection{\mydataset Dataset}
We conduct our experiments on a custom dataset, \mydataset. Unlike existing object-centric datasets such as TUM RGB-D~\cite{tum} or single-user datasets such as ScanNet~\cite{dai2017scannet}, Replica-Share is specifically designed for map sharing and evaluating view extrapolation. 
To the best of our knowledge, no publicly available dataset exists that is tailored for this purpose.
\mydataset is recorded using the Replica simulator~\cite{replica} and spans eight diverse indoor scenes. For each scene,  we collect 2,000 observations along the trajectories of three different contributors.
To save bandwidth, only one out of every 20 observations is uploaded to the server, while all 2,000 observations are retained for evaluating the view interpolation performance of the global map. 
For view extrapolation evaluation, we generate additional ground-truth data by uniformly sampling 100 positions per scene and pairing each position with 10 random rotations, producing 1,000 novel views per scene, most of which are extrapolated.

\subsection{Metrics}
We evaluate PSNR, LPIPS, SSIM, and Depth L1 for both interpolated and extrapolated views under \mysystem, HAC, and 3DGS. Additionally, we report the size of the data transmitted from the server to the user. 
For seen areas, this includes the bitstream of the map increment $F_{Inc}$ and its corresponding decoder, as shown in Fig.~\ref{fig:overview}. For unseen areas, it includes the bitstream of the full map $F_{anc}$, its corresponding decoder, and additional anchor positions $V$. 
We calculate the size of the accumulated transmission data across all stages as the map size. The compression ratio is then determined by comparing this map size to that of the original 3DGS.

\subsection{Results} 
\noindent \textbf{Overall Performance.}
We compare the performance of \mysystem, HAC~\cite{hac}, and the original 3DGS~\cite{3DGS} on both extrapolated and interpolated views, as shown in Tab.~\ref{tab:overall_extra} and Tab.~\ref{tab:overall_inter}, with a particular focus on the performance of the third stage. 
For HAC and 3DGS, the full map is transmitted across all three stages. In contrast, \mysystem transmits the full map only for unseen areas, and updates seen regions using compact map increment features along with a corresponding decoder.
The results reveal that \mysystem achieves significantly higher fidelity and compression ratios compared to the other methods across all scenes in view extrapolation. 
Taking the scene Office2 as an example, \mysystem achieves a PSNR that is 4.24 dB higher than HAC and 3.74 dB higher than 3DGS, representing improvements of 15\% and 13\%, respectively. Additionally, the transmission overhead of \mysystem is reduced by 36\% compared to HAC, with a compression ratio of 86 times relative to the original 3DGS. 
Across all eight scenarios, the PSNR, LPIPS, SSIM, and Depth L1 of \mysystem improve by 11\%, 22\%, 2.8\%, and 74\% compared to the state-of-the-art method HAC, respectively, while the transmission overhead is reduced by 36\%.   
The improvements in compression efficiency are attributed to two key design choices. First, the initial global map is transmitted in an implicit anchor-based representation, which is significantly more compact than explicitly storing all Gaussians, as in 3DGS. Second, for map updates, we transmit only map increment information, effectively reusing the client-side cached map to eliminate redundant data transfer, offering a more efficient update mechanism than HAC.
For interpolated views, we observed that \mysystem performs similarly to HAC, while 3DGS excels. This is primarily because both \mysystem and HAC employ compression, which inevitably introduces some degradation, whereas 3DGS uses a full, uncompressed representation. The high performance of 3DGS, however, comes at the cost of a significantly larger map size.
On extrapolated views, \mysystem outperforms both baselines, which can be attributed to the proposed virtual-image-based map enhancement module. 
Since interpolated views are already accurate, we did not observe significant improvements in those cases.  
 
\begin{figure*}[!t]
\centering
    \begin{minipage}[t]{0.24\linewidth}
       \centering 
        \includegraphics[width=0.95\linewidth]{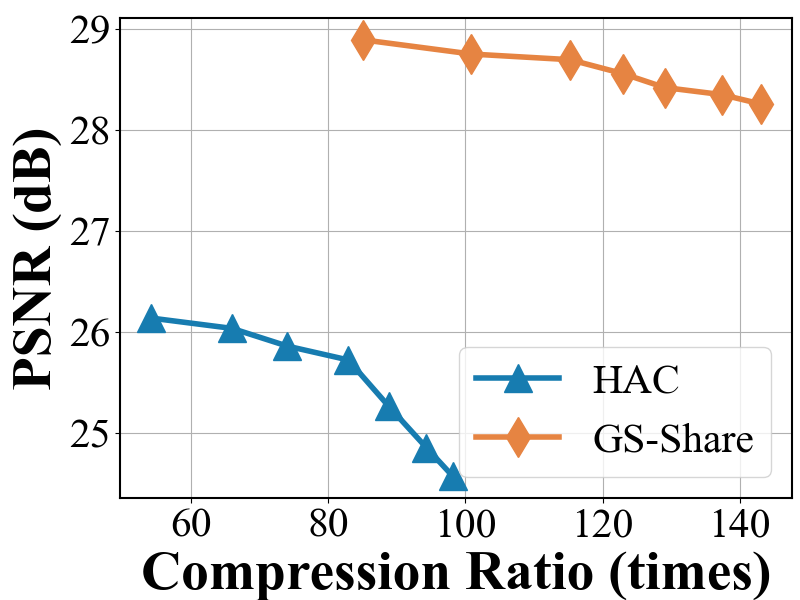}\\ 
    \caption{Performance comparison under different compression ratios on extrapolated views.} \label{fig:1_hac} 
    \end{minipage}%
    \hspace{0.5em}
    \begin{minipage}[t]{0.24\linewidth}
        \centering 
        \includegraphics[width=0.95\linewidth]{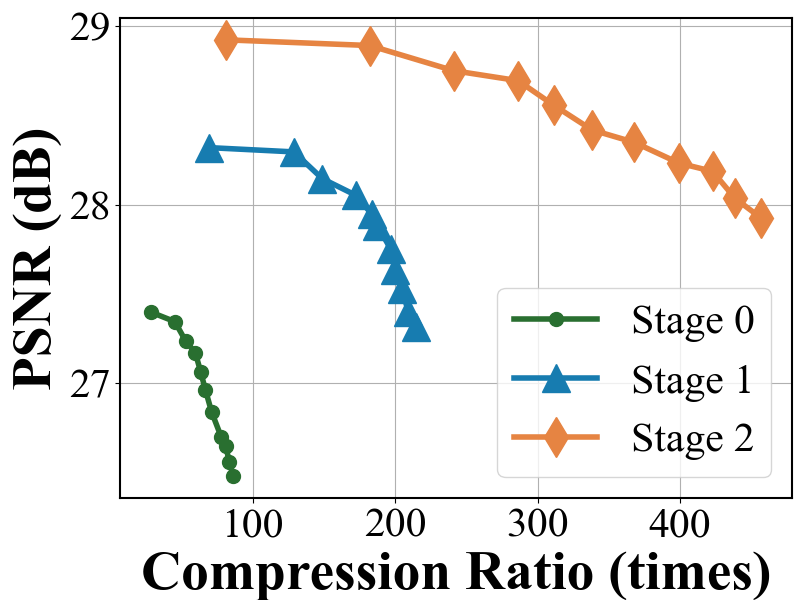}\\ 
    \caption{Performance comparison of \mysystem across different stages on extrapolated views.} \label{fig:0_stage}  
    \end{minipage}%
    \hspace{0.5em}
    \begin{minipage}[t]{0.24\linewidth}
        \centering  
        \includegraphics[width=0.95\linewidth]{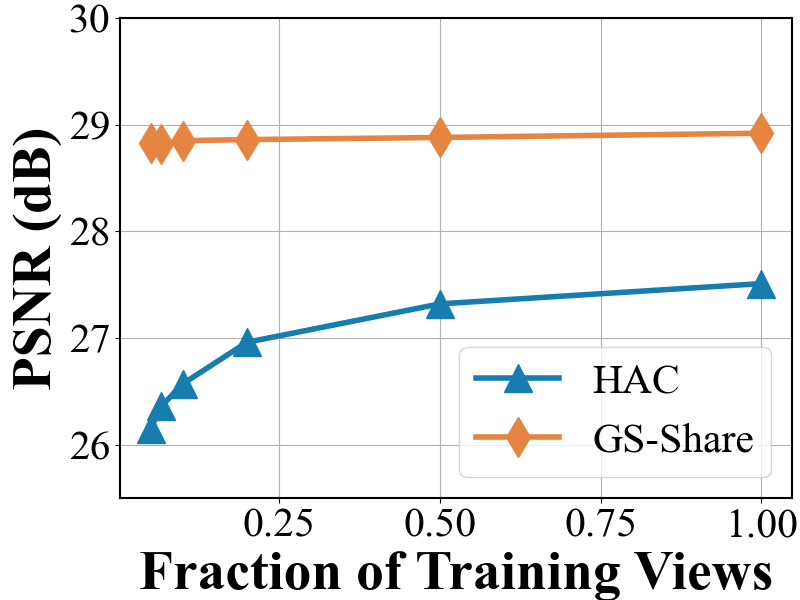}\\ 
    \caption{Impact of fraction of training views, evaluated on extrapolated views.}  \label{fig:4_number}
    \end{minipage}%
    \hspace{0.5em}
    \begin{minipage}[t]{0.24\linewidth}
        \centering  
        \includegraphics[width=0.95\linewidth]{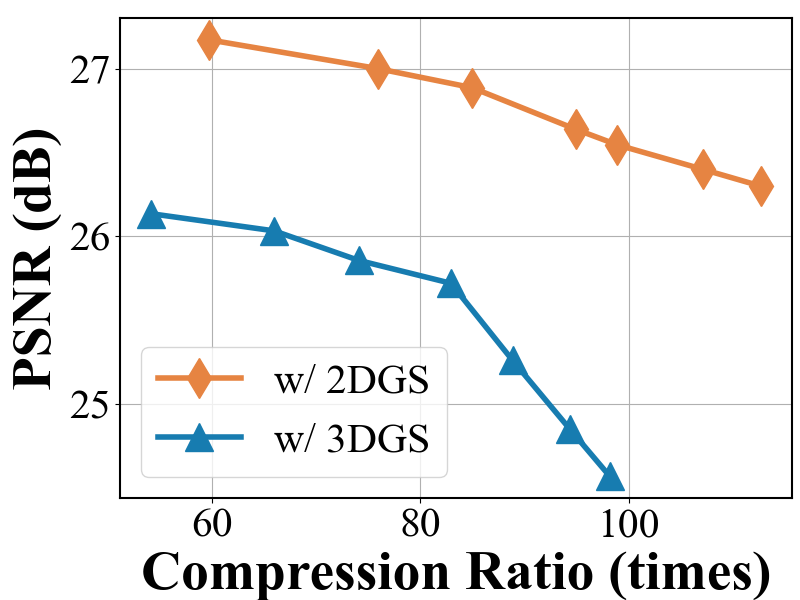}\\ 
    \caption{Impact of Gaussian representation, evaluated on extrapolated views.  }  \label{fig:2_2DGS}
    \end{minipage}  
    \hspace{0.5em}
    \vspace{-1em}
\end{figure*}

\noindent \textbf{Performance across Different Compression Ratios.} 
To illustrate the impact of compression ratios of HAC and \mysystem, we conducted experiments at different compression ratios on \mydataset Room0. As shown in Fig.~\ref{fig:1_hac}, \mysystem maintains high fidelity even at higher compression ratios, for instance, at a compression ratio of 143 times, the PSNR is 28.25 dB.
In contrast, for HAC, the results indicate a significant degradation in the PSNR curve when the compression ratio is approximately 82 times.

\noindent \textbf{Performance across Different Stages.} 
Fig.~\ref{fig:0_stage} illustrates the system's performance across different stages and compression ratios. Experiments are conducted on the \mydataset Room0. We control the compression ratio by setting $\lambda_{q}$ from 0.0005 to 0.0205 in increments of 0.002. 
Please note that in this experiment, the compression rate is computed on a per-stage basis to enable clearer comparison, rather than being aggregated across all stages. 
As shown in Fig.~\ref{fig:0_stage}, we observe that within the same stage, PSNR declines rapidly as the compression ratio increases. For example, as the compression ratio increases from 28 to 86 times in stage 0, from 69 to 214 times in stage 1, and from 81 to 457 times in stage 2, the PSNR decreases by 0.92 dB, 1.01 dB, and 1.00 dB, respectively. 
However, the highest achieved PSNR steadily increases from stage 0 to stage 2, indicating that additional observations contribute to improved map quality. 
In addition to PSNR, the compression ratio also increases with the stage ID,  demonstrating the effectiveness of the map increment.

\begin{table}[b]\centering  
\setlength\tabcolsep{0.6pt} 
\renewcommand\arraystretch{1} 
\caption{\label{tab:moduleAblation} Ablation study of three main components. 
} 
\vspace{-4pt}
\begin{tabular}{cccc||c|c|c|c|c}
\hline
\textbf{Base} & \textbf{VIRT} & \textbf{ANCR} & \textbf{INCR} & \makecell[c]{{PSNR} \\ {[dB]}$\uparrow$} & {LPIPS$\downarrow$} & {SSIM$\uparrow$} & \makecell[c]{{Depth L1} \\ {[cm]$\downarrow$}} & \makecell[c]{{Size}  \\ {[MB]$\downarrow$} }   \\
\hline \hline
\checkmark &  &  &  & 26.10 & 0.184 & 0.898 & 2.80 &  5.81 \\
\checkmark & \checkmark &  &  & 28.17 & 0.168 & 0.928 & 1.75 & 6.06\\
\checkmark & \checkmark & \checkmark &  & 28.58 & 0.145 & 0.935 & 1.08 & 5.56 \\
\checkmark & \checkmark & \checkmark & \checkmark & 28.83 & 0.145 & 0.937 & 1.08 &  3.72 \\
\hline
\end{tabular} 
\end{table}

\begin{table}[b]\centering  
\setlength\tabcolsep{0.1pt} 
\renewcommand\arraystretch{1} 
\caption{\label{tab:smallAblation} Ablation study of additional techniques.} 
\vspace{-4pt}
\begin{tabular}{c||c|c|c|c|c}
\hline
\textbf{Experiment} & \makecell[c]{{PSNR} \\ {[dB]}$\uparrow$} & {LPIPS$\downarrow$} & {SSIM$\uparrow$} & \makecell[c]{{Depth L1} \\ {[cm]$\downarrow$}} & \makecell[c]{{Size}  \\ {[MB]$\downarrow$} }   \\
\hline \hline
\mysystem & 27.30 & 0.155 & 0.917 & 1.28 &  2.35\\
w/o Edge Filtering & 27.00 & 0.160 & 0.914 & 1.43 & 2.35 \\
w/o Hole Filling & 25.97 & 0.173 & 0.907 & 1.32 & 2.36 \\
w/o Confidence Prediction & 26.88 & 0.163 & 0.915 & 1.38 & 2.33 \\ 
\hline
\end{tabular}  
\end{table}

\begin{figure*}[!t]
	\centering
	\includegraphics[width = 0.97\linewidth]{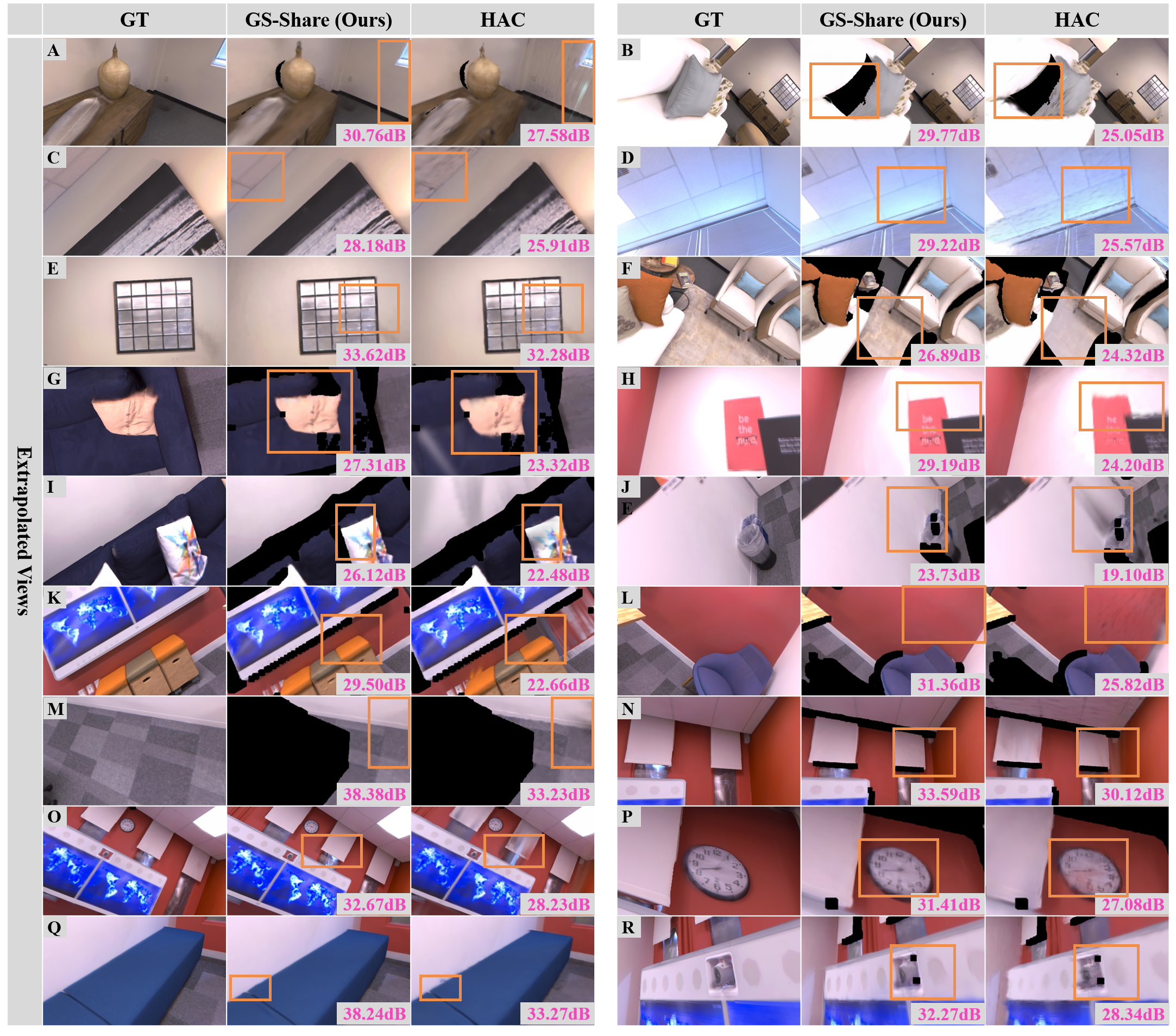} 
    \vspace{-6pt}
	\caption{Visualization results of \mysystem and HAC~\cite{hac}. We present the ground truth alongside the rendering results of \mysystem and HAC. The PSNR values are labeled in the bottom right corner of each image.}
	\label{fig:visual_hac} 
    \vspace{-1em}
\end{figure*}

\noindent \textbf{Ablation Study of the Main Components.} 
As shown in Tab.~\ref{tab:moduleAblation}, we conduct an ablation study at stage 2 on three main components: virtual-image-based map enhancement (VIRT), modified anchor-based representation (ANCR), and the representation of map increment (INCR). All performances are reported for extrapolated views.
The ``Base'' setting refers to HAC. 
The results indicate that the use of VIRT improves the PSNR by 2.07 dB over the baseline, while LPIPS, SSIM, and Depth L1 also show improvements of 0.016, 0.030, and 1.05 cm, respectively. 
Building upon VIRT, the use of our modified anchor-based representation further improves the PSNR by 0.41 dB, LPIPS by 0.023, and Depth L1 by 0.67 cm, while also resulting in a slight map size reduction of 8.3\%.  
Finally, the introduction of INCR improves the PSNR by an additional 0.25 dB and, more importantly, reduces the map size from 5.56 MB to 3.72 MB, representing a 33\% reduction.
To further quantify the effect of the number of input views, we conduct an extensive experiment. Specifically, we control the fraction of training views -- ranging from 20× downsampling to using the full set -- and show the PSNR values for extrapolated views in Fig.~\ref{fig:4_number}.  
These results show that \mysystem consistently improves performance on extrapolated views under various settings, validating its effectiveness. 

\noindent \textbf{Ablation Study of Additional Techniques.} Except for the main components, we also conduct an ablation study on several additional techniques, including edge filtering discussed in Sec.~\ref{subsec:anchor}, hole filling, and confidence prediction discussed in Sec.~\ref{subsec:virtual}. 
Results for stage 0 are reported to better elucidate the effects of these modules. 
As shown in Tab.~\ref{tab:smallAblation}, the inclusion of edge filtering increases the PSNR by 0.3 dB and reduces Depth L1 by 0.15 cm, as normal estimation is closely related to the object's geometry. 
Additionally, both the hole filling and confidence prediction modules show improvements in PSNR, LPIPS, SSIM, and Depth L1, as they are designed to enhance the quality of the pseudo ground truth. 
Finally, all techniques have a negligible impact on the map size, which is in line with expectations.

\noindent \textbf{Ablation Study of Different Gaussian Representations.} 
Next, we examine the effect of Gaussian representation under varying compression ratios. To ensure a fair comparison, these experiments are conducted without virtual image enhancement or incremental map updates. 
As shown in Fig.~\ref{fig:2_2DGS}, the proposed ANCR module using modified 2D Gaussians outperforms its 3D Gaussian counterpart in terms of PSNR.  
At a compression ratio of 98 times, the 2D Gaussian variant achieves a PSNR of 26.55 dB, compared to 24.57 dB for the 3D Gaussian representation. Moreover, the performance gap widens as the compression ratio increases, demonstrating that ANCR is better suited for view extrapolation under compact Gaussian splatting constraints.

\noindent \textbf{Visualization Results of \mysystem and HAC.} We visualize the qualitative results of HAC~\cite{hac} and \mysystem on extrapolated views at stage 2 in Fig.~\ref{fig:visual_hac}. Regions with significant differences are highlighted with orange boxes, and the PSNR is labeled in the bottom right corner of each image. 
We observe that in areas with insufficient observations, view extrapolation under HAC often leads to several negative effects, such as artifacts, distortion, and incorrectly rendered colors and textures. In contrast, the rendered images from \mysystem exhibit higher fidelity and substantially reduce these issues, showcasing the effectiveness of the designed modules.

\begin{figure*}[!t]
	\centering
	\includegraphics[width = 0.97\linewidth]{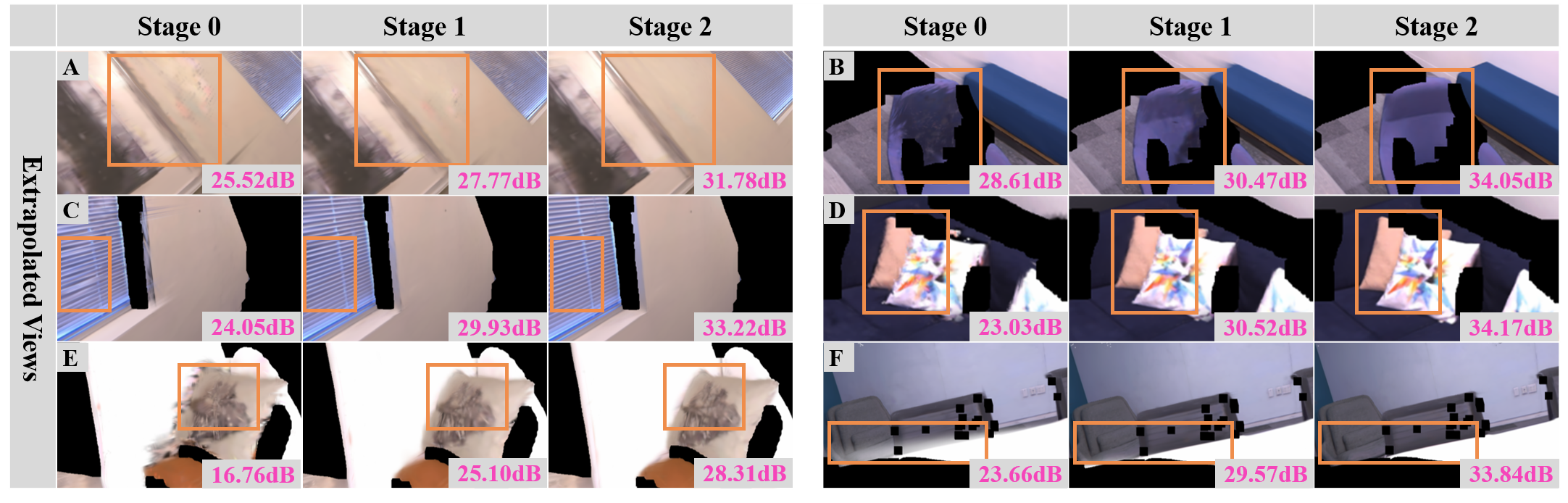} 
    \vspace{-6pt}
	\caption{Visualization results of \mysystem at different stages. The PSNR values are labeled in the bottom right corner of each image.}
	\label{fig:visual_stage} 
    \vspace{-1em}
\end{figure*}

\noindent \textbf{Visualization Results of \mysystem at Different Stages.} We present the visualization results of \mysystem across multiple stages in Fig.~\ref{fig:visual_stage}. Regions with significant differences are highlighted with orange boxes. As the stage progresses, the reconstruction quality consistently improves. This is attributed to the increasing amount of input data, which provides richer information for reconstruction.
This highlights the importance of timely map updates in map-sharing systems. 
The improvement is especially noticeable in areas with fine textures, such as object boundaries and detailed structures like window blinds.

\section{Conclusion}
\label{sec:conclusions}
In this work, we propose \mysystem, the first incremental map-sharing system based on Gaussian splatting. \mysystem enables high-fidelity rendering, efficient map sharing, and continuous map updating with a compact representation.
Our experiments demonstrate that \mysystem outperforms state-of-the-art methods in both view extrapolation and transmission overhead.   
Moving forward, we will continue to explore more practical challenges that a real-world map-sharing system may encounter, including the integration of various types of sensors, each with differing qualities.

\section{ACKNOWLEDGMENTS}  
This work was supported by the National Natural Science Foundation of China (No. 62332016) and the Key Research Program of Frontier Sciences, CAS (No. ZDBS-LY-JSC001). 

\bibliographystyle{eg-alpha-doi} 
\bibliography{Main}     
\end{document}